\documentclass[twoside,12pt]{article}  
\usepackage{CJK}   
\usepackage{indentfirst}  
\usepackage{bm}    
\usepackage{graphicx}  
\usepackage{enumitem}
\usepackage{amssymb}
\usepackage{pifont}
\usepackage{latexsym}
\usepackage{amsmath}
\usepackage{subfigure}
\usepackage{epsfig}

\usepackage{color}
\usepackage{ulem}

\begin{document}    

\begin{CJK*}{GBK}{song}  

\thispagestyle{empty} \vspace*{0.8cm}\hbox
to\textwidth{\vbox{\hfill\huge\sf \hfill}}
\par\noindent\rule[3mm]{\textwidth}{0.2pt}\hspace*{-\textwidth}\noindent
\rule[2.5mm]{\textwidth}{0.2pt}


\begin{center}
\LARGE\bf Threshold for the Outbreak of Cascading Failures in Degree-degree Uncorrelated Networks   
\end{center}

\footnotetext{\hspace*{-.45cm}\footnotesize $^*$
}
\footnotetext{\hspace*{-.45cm}\footnotesize $^\dag$}

\begin{center}
\rm Junbiao Liu$^{\rm a}$, \ \ Xinyu Jin$^{\rm a}$, \ \ Lurong Jiang$^{\rm b \rm a}$, \ \ Yongxiang Xia$^{\rm a}$, \ \ Bo Ouyang$^{\rm c}$, \ \ Fang Dong$^{\rm d}$, \ \ Yicong Lang$^{\rm a}$, \ \ Wenping Zhang$^{\rm a}$
\end{center}

\begin{center}
\begin{footnotesize} \sl
$ {College of Information Science and Electronic Engineering, Zhejiang University, Hangzhou, China.}^{\rm a)}$ \\   
$ {School of Information Science and Technology, Zhejiang Sci-Tech University, Hangzhou, China.}^{\rm b)}$ \\
$ {College of Electrical and Information Engineering, Hunan University, Changsha, China.}^{\rm c)}$ \\   
$ {School of Information and Electric Engineering, Zhejiang University City College, Hangzhou, China.}^{\rm d)}$ \\   
\end{footnotesize}
\end{center}

\begin{center}
\footnotesize 
\end{center}

\vspace*{2mm}

\begin{center}
\begin{minipage}{15.5cm}
\parindent 20pt\footnotesize

In complex networks, the failure of one or very few nodes may cause cascading failures. When this dynamical process stops in steady state, the size of the giant component formed by remaining un-failed nodes can be used to measure the severity of cascading failures, which is critically important for estimating the robustness of networks. In this paper, we provide a cascade of overload failure model with local load sharing mechanism, and then explore the threshold of node capacity when the large-scale cascading failures happen and un-failed nodes in steady state cannot connect to each other to form a large connected sub-network.
We get the theoretical derivation of this threshold in degree-degree uncorrelated networks, and validate the effectiveness of this method in simulation. This threshold provide us a guidance to improve the network robustness under the premise of limited capacity resource when creating a network and assigning load. Therefore, this threshold is useful and important to analyze the robustness of networks.
\end{minipage}
\end{center}

\begin{center}
\begin{minipage}{15.5cm}
\begin{minipage}[t]{2.3cm}{\bf Keywords: }\end{minipage}
\begin{minipage}[t]{13.1cm}
complex networks, cascading failures, load absorbing nodes, the giant component
\end{minipage}\par\vglue8pt
{\bf PACS: }
89.75.-k, 05.70.Jk, 05.10.-a
\end{minipage}
\end{center}

\section{Introduction}  



Cascading failures are a sort of phenomena that a random failure or intentional attack on one or a few nodes lead to serve chain reaction in the networks, which can cause collapse of a large fraction of nodes in the network. It is widely found in many real-world networks, such as power transmission \cite{sachtjen2000disturbances,kinney2005modeling}, communication \cite{cohen2000resilience}, economic \cite{huang2013cascading}, and biological \cite{borrvall2000biodiversity} networks. A real example of cascading failures is the well-known Northeast Blackout in 2003 \cite{2003Blackout}. In this case, the outage of a generator led to a serve chain reaction of power blackout, which affected approximately 50 million people in North America and caused financial losses of about \$6 billion.

In order to understand the essential mechanism of cascading failures, a number of cascading models have been proposed, such as betweenness-based model \cite{motter2002cascade,wang2008attack,xia2013EPL,xia2010PhysicaA}, sand-pile model \cite{goh2003sandpile,hoore2013critical}, and fiber-bundle model \cite{moreno2002instability,kim2005universality}, etc. In these models, a node fails when its load exceeds its capacity. The failure of this node leads to the redistribution of load in network, and can cause collapse of a large fraction of the network. Therefore, the cascading failure process of these models is highly depended on the relation between load and capacity. 

When the dynamical process of cascading failures terminates in steady state, the network breaks into several connected sub-networks formed by un-failed nodes. The size of the largest connected sub-network (i.e., the giant component) can be used to measure the severity of cascading failures, which is critically important for estimating the robustness of networks \cite{PhysRevLett.106.048701,ganesan2013size}. It is intuitive that when the capacity increases (or the load decreases), the size of cascading failures in network reduces, which is confirmed by all of those cascading failure models above. In fiber bundle and sandpile model, with the capacity under a critical value (or load above a critical value), the giant component disappears, or above which the size of the giant component dramatically rises up \cite{moreno2002instability,lee2004sandpile}.
This critical value is a very important feature to measure the robustness of networks. However, to our knowledge, there is little research on quantitative analysis of the relation between node capacity and this threshold for the break out of cascading failures at present.

In this paper, we provide a cascade of overload failure model with local load sharing mechanism, then explore the threshold of node capacity when the large-scale cascading failures happen and there does not exist a large connect sub-network formed by un-failed nodes.


\section{Model}  
Here we provide a cascade of overload failure model with local load sharing mechanism. In this model, the statuses of nodes are divided into two categories: the \textbf{un-failed} and \textbf{failed}. We assume that all nodes in networks are un-failed at the beginning. A node fails if its load exceeds its capacity (i.e., overload). 
When a node fails to work, this node is considered transferring a fixed positive load $\Delta$ to each of its un-failed neighbors and being separated from the giant component \cite{dobson2005loading,sansavini2009deterministic}. When the cascading failures terminate, only those nodes in the giant component are supposed to work. It is natural to assume that the capacity $C_v$ of a node $v$ is proportional to its initial load $L_v$ \cite{motter2002cascade,lehmann2010stochastic} as

\begin{equation}\label{eq:cap}
C_v=(1+\alpha)L_v           \; ,
\end{equation}

\noindent where the constant $\alpha$ is the tolerance parameter. The initial load of each node is randomly distributed following a uniform distribution on the interval $[ L_{min}, L_{max} ]$. For simplicity, we set $L_{min} = 0$ and $L_{max} = 1$ \cite{dobson2005loading,sansavini2009deterministic}. Then for arbitrary node $v$, the cumulative distribution function of $L_v$ is

\begin{equation}
\label{eq:load_cdf}
P\{ L_v<l \} \triangleq \varphi(l) =
\begin{cases}
0 & \text{$l \leq 0$} \\
l & \text{$0 < l \leq 1$} \\
1 & \text{$l > 1$}
\end{cases} \; .
\end{equation}

The numerical process of cascading failures with local load sharing mechanism is summarized as follows:

\begin{enumerate}
  \item Initialization. Generate a degree-degree uncorrelated network with $N$ nodes. Assume all nodes are un-failed. The load of each node is uniformly distributed in $[L_{min}, L_{max}]$, where we let $L_{min} = 0$ and $L_{max} = 1$. The capacities of nodes are determined by Eq (\ref{eq:cap}).
  \item Beginning. Choose very few node randomly in the network, and set them as failed.
  \item\label{label:load_redistribution} Load redistribution. In each round, each node which is failed in the last round transfers a fixed positive load $\Delta$ to each of its un-failed neighbors. An un-failed node turns to be failed if it overloads in this round.  
  \item Halt. Repeat step \ref{label:load_redistribution} if there exist overloaded nodes in the network; otherwise, the process halts. Finally, only the un-failed nodes in the giant component are supposed to work.
\end{enumerate}

%



\section{Analysis}

\subsection{Description of critical conditions}     
When the cascading failure process ends, the un-failed nodes in the network form several connected sub-networks. The fraction of giant component, which is the relative size of the largest connected sub-network, can be used as a measure of the network performance against cascading failures. Fig. 1 shows the fraction $G$ of the giant component as a function of tolerance parameter $\alpha$ in ER random networks \cite{erd6s1960evolution} and BA scale-free networks \cite{barabasi1999emergence}. For each $G$ with different average degree $\langle k \rangle$, there exists a critical tolerance parameter $\alpha_c$, with $\alpha$ under which the giant component disappears, and over which $G$ increases dramatically and approaches to $1$ finally.
When $\alpha < \alpha_c$, $G$ approximately equals to $0$, which indicates large-scale cascading failures occur and network breaks into extremely small clusters. There are two conditions to ensure the $G$ approximately equals to $0$ when $\alpha < \alpha_c$:

I. Large-scale cascading failures occur in the network, i.e., the failed nodes connect to each other to form a large connected sub-network.

II. When the dynamical process of cascading failures terminates in steady state, there does not exist a large connected sub-network formed by un-failed nodes. 

In the rest of this section, we give a theoretical derivation of $\alpha_c$ with these two conditions above.


\vspace*{4mm}
\begin{figure}
\centering
\subfigure[ER random networks.]{
\resizebox{0.45\textwidth}{!}{
  \includegraphics{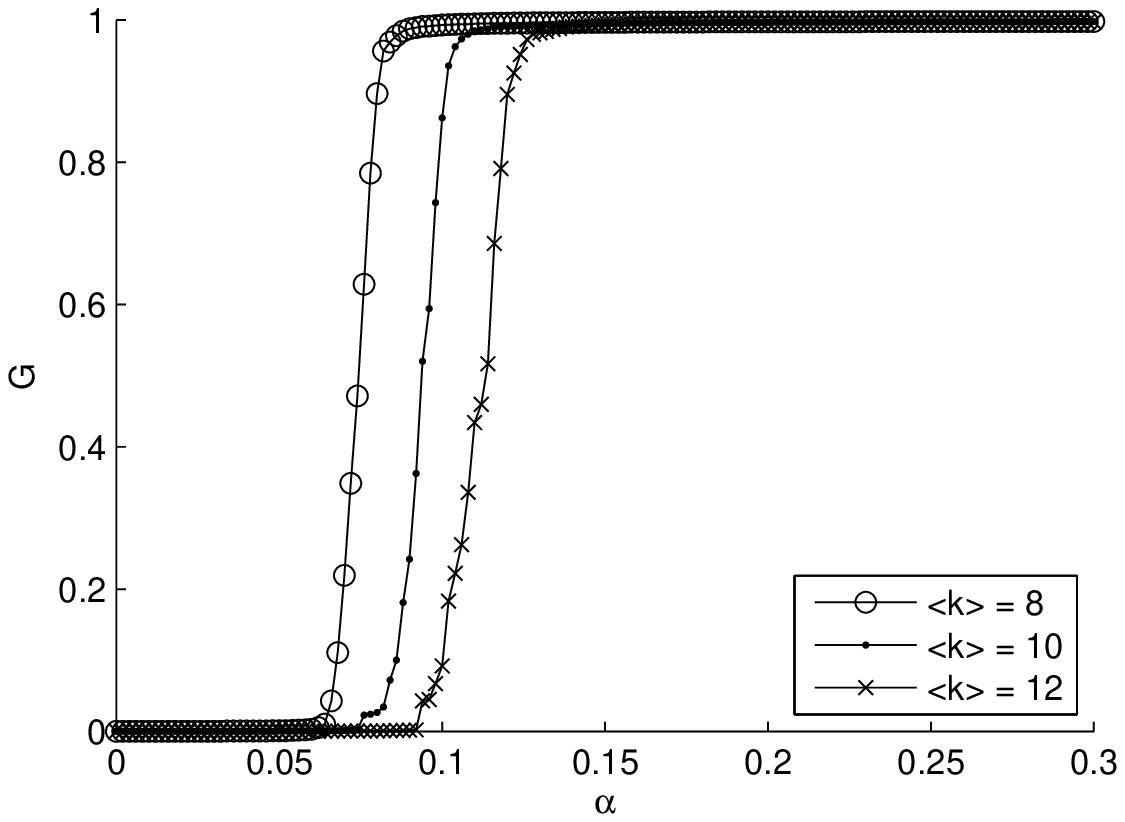}
}
}
\subfigure[BA scale-free networks.]{
\resizebox{0.45\textwidth}{!}{
  \includegraphics{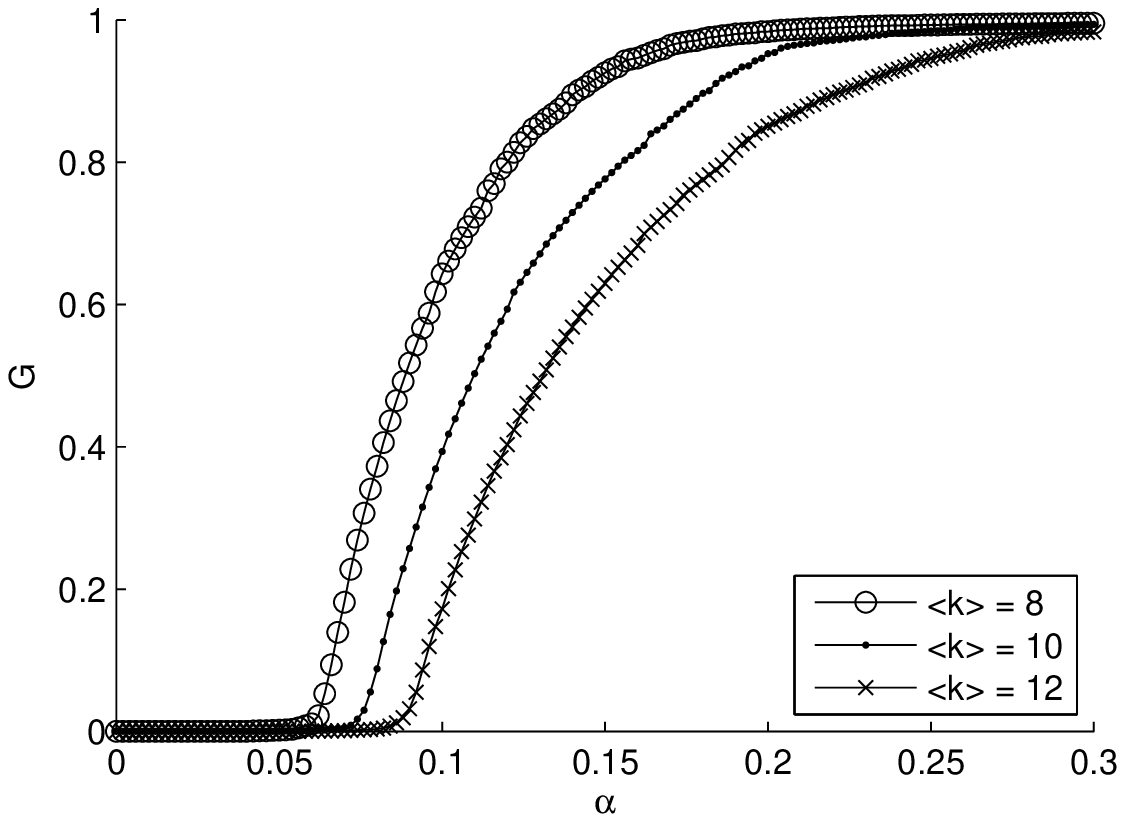}
}
}
\parbox{15.5cm}{\small{\bf Fig. 1}
The fraction $G$ of giant component changes with increase of tolerance parameter $\alpha$ in ER random and BA scale-free networks.}
\label{fig:giant_component_changes}
\end{figure}

\subsection{Large-scale cascading failures occur in the network}     

First, let us consider condition I that large-scale cascading failures occur in the network.  This condition equals to the situation that the failed nodes connect to each other to form a large connected sub-network. On average, each node fails and causes more than one of its neighbors to fail when large-scale cascading failures happen. Consider a node $v$ of degree $k_v$, and any of its neighbor $w$. The probability that node $v$ fails and causes node $w$ to fail is

\begin{equation}
\begin{split}
P \left\{ L_w + \Delta>(1+\alpha)L_w \right\} &=P \left\{ L_w<\frac{\Delta}{\alpha} \right\}
=\varphi \left( \frac{\Delta}{\alpha} \right)           \; ,
\end{split}
\end{equation}

\noindent where $L_w$ is the initial load of node $w$.

The cascading failures of nodes are a sort of site percolation process. Because of the locally tree-like approximation in percolation of degree-degree uncorrelated networks, the probability that node $v$ fails and causes $m$ of its $k_v-1$ neighbors (subtract the node which causes node $v$ to fail) to fail is

\begin{equation}
\binom{k_v-1}{m} \varphi \left( \frac{\Delta}{\alpha} \right)^m \left( 1-\varphi \left(\frac{\Delta}{\alpha}\right) \right)^{k_v-1-m}   \; .
\end{equation}

Thus, on average, the node $v$ fails and causes its

\begin{equation}
\label{eq:avg_f_neibs}
\begin{split}
&\sum_{m=1}^{k_v-1} m \binom{k_v-1}{m} \varphi\left( \frac{\Delta}{\alpha} \right) ^m \left( 1-\varphi\left( \frac{\Delta}{\alpha}\right)\right)^{k_v-1-m} \\
=&\sum_{m=1}^{k_v-1} m \binom{k_v-1}{m} \varphi\left( \frac{\Delta}{\alpha} \right) ^m \left( 1-\varphi\left( \frac{\Delta}{\alpha}\right)\right)^{k_v-1-m} \\
=&\sum_{m=1}^{k_v-1} (k_v-1) \binom{k_v-2}{m-1} \varphi\left( \frac{\Delta}{\alpha} \right)^m\left(1-\varphi\left(\frac{\Delta}{\alpha}\right)\right)^{k_v-1-m} \\
=&(k_v-1)\varphi\left( \frac{\Delta}{\alpha} \right)
\end{split}
\end{equation}

\noindent neighbors to fail.

Denote $p_k$ as the probability that a randomly-picked node has degree $k$. According to Eq (\ref{eq:avg_f_neibs}), the failure of arbitrary node in the network causes its
$\sum_{k=0}^{\infty} p_k \left( k-1 \right) \varphi \left( \frac{\Delta}{\alpha} \right)= \left( \langle k \rangle - 1 \right) \varphi \left( \frac{\Delta}{\alpha} \right)$
neighbors to fail. When large-scale cascading failures occur in the network, the failure of arbitrary node causes more than one of its neighbors to fail, that is

\begin{equation}
\label{eq:ineq_alpha_l}
\left( \langle k \rangle - 1 \right) \varphi \left( \frac{\Delta}{\alpha} \right) \geq 1            \; .
\end{equation}

Thus, for condition I that large-scale cascading failures occur in the network, we have

\begin{equation}
\label{eq:alpha_l_cond1}
\alpha \leq \left( \langle k \rangle - 1 \right) \Delta     \; .
\end{equation}

It is worth noting that there is not any specific assumption on degree distribution for the result above.

\subsection{There does not exist a large connected sub-network formed by un-failed nodes}

According to the local load sharing mechanism in our model, there may exist some nodes who do not propagate cascading failures. These nodes never fail with large tolerance parameter $\alpha$, or their failure will not affect other nodes as all of their neighbors are already failed. We call these nodes as \textbf{absorbing nodes}. Before handling with condition II, we pay close attention to these absorbing nodes. Let us consider the situation that a node $v$ of degree $k_v$ does not fail after $k_v -1$ of its neighbors fail and share loads $(k_v -1) \Delta$ to it. Consequently, no matter node $v$ fails or not, it can not affect other nodes any more. In this situation, node $v$ have the ability of absorbing the loads of all its neighbors by itself, thus we call this node as the \textbf{independent absorbing node}. The capacity and initial load of an independent absorbing node $v$ meet the following relation as

\begin{equation}
\label{eq:independent_absorbing_nodes_def}
L_v+ \left( k_v - 1\right) \Delta < \left( 1+\alpha \right) L_v \; .
\end{equation}

But this is not the only reason that a node happens to be an absorbing node. There is another situation for a node to be an absorbing node whose load does not satisfy the Eq (\ref{eq:independent_absorbing_nodes_def}). In this situation, considering a node $v$, $m$ of its neighbors are absorbing nodes, thus we only need  $n_v > k_v - m - 1 $ to make node $v$ an absorbing node. We call this kind of nodes as \textbf{dependent absorbing nodes} because their abilities of absorbing loads depend on the other absorbing nodes of their neighbors. Dependent absorbing nodes can also prevent cascading failures locally.


Fig. 2 gives two examples of absorbing nodes. In Fig. 2(a), the un-failed node 1 has five neighbors and four of them (node 2, 3, 4, 5) are failed. Assume that if node 6 does not fail in the following procedure of cascading failure, then node 1 does not fail either. Or in another case that node 6 fails and transfers a fixed value of load $\Delta$ to node 1. In the latter case, no matter node 1 fails or not, it can not affect any other node. Thus, node 1 is an independent absorbing node. In Fig. 2(b), we assume that the node 7 is not an independent absorbing node. It happens to fail if more than two of its four neighbors fail and transfer loads to it. Assume node 8 is an absorbing node, thus it never fail. Consequently, no matter node 7 fails or not, it can not share load to any un-failed neighbor. Therefore, node 7 in Fig. 2(b) comes to be a dependent absorbing node because its ability of absorbing loads depends on the absorbing node 8.

\vspace*{4mm}

\begin{figure}
\setcounter{subfigure}{0}
\centering
\subfigure[The independent absorbing nodes.]{
\resizebox{0.35\textwidth}{!}{
  \includegraphics{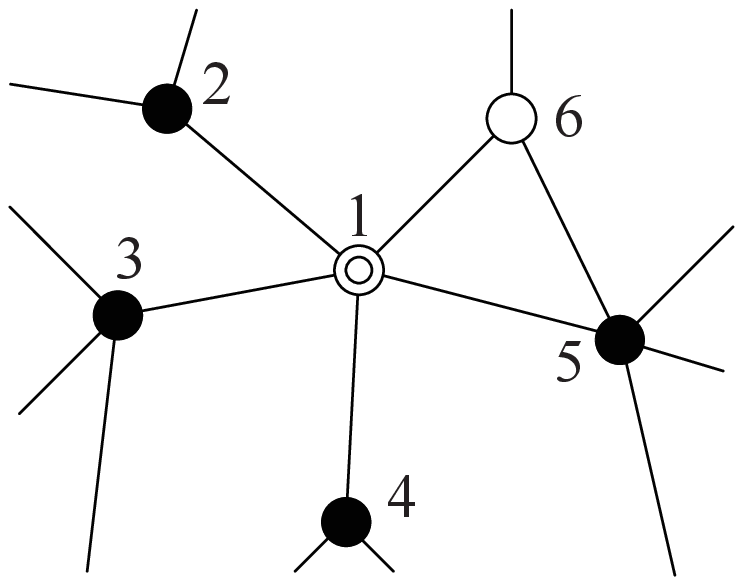}
}
}
\subfigure[The dependent absorbing nodes.]{
\resizebox{0.35\textwidth}{!}{
  \includegraphics{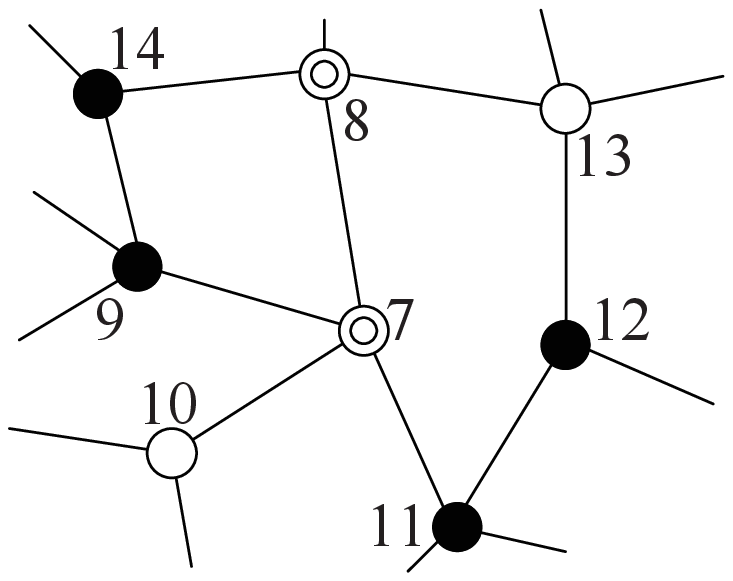}
}
}
\parbox{15.5cm}{\small{\bf Fig. 2}
Two examples of the absorbing nodes. In the procedure of cascading failures, an un-failed node ($\circ$) may turn to be a failed node ($\bullet$) when it overloads, or does not fail as it is an absorbing node ($\circledcirc$).}
\label{fig:absorbing_nodes_demo}
\end{figure}

Then we consider condition II that there does not exist a large connected sub-network formed by un-failed nodes.  We have explained that absorbing nodes survive and keep un-failed after cascading failures. With increase of $\alpha$, each node has higher probability to be assigned with more capacity and be an absorbing node on average. If there are a large number of absorbing nodes exist, they may connect to each other to form a large connected sub-network which makes $G$ greater than 0.
We have divided absorbing nodes into independent absorbing nodes and dependent absorbing nodes. 
Further more, we call an absorbing node as an \textbf{\textit{m}-absorbing node} ($m = 0, 1, 2, ... , k$), if and only if $m$ of its neighbors are absorbing nodes.
We assume that an independent absorbing nodes is not equivalent to 0-absorbing node. For an interdependent node, if it has $m$ absorbing nodes of its neighbors, we still treat this node as an $m$-absorbing node.

Then, we let the probability that a node of degree $k$ happens to be an absorbing node as $a_k$. The probability that node $v$ happens to be an $m$-absorbing node is $a_k^{(m)}$. Then we have

\begin{equation}
\label{eq:absorb_prob_decomposition}
a_k = \sum_{m=0}^{k} a_k^{(m)}  \; .
\end{equation}

In a degree-degree uncorrelated network, consider a node $v$ of degree $k$ and any of its neighbor $w$. The probability of node $w$ with degree $j$ is $\frac{j p_j}{\sum_iip_i} = \frac{j p_j}{\langle k \rangle}$. Node $w$ happens to be an absorbing node with the probability $a_j$. Then, the probability of node $v$'s arbitrary neighbor being an absorbing node is

\begin{equation}
\label{eq:sigma_a}
\sigma_a = \sum_{j} \frac{j p_j a_j}{\langle k \rangle}         \; .
\end{equation}

According to Eq. (\ref{eq:sigma_a}), the probability that exactly $m$ of node $v$'s neighbors are absorbing nodes is

\begin{equation}
\binom{k}{m} \sigma_a^m (1-\sigma_a)^{k-m}          \; .
\end{equation}

Node $v$ is an $m$-absorbing node if it stays un-failed with $k-m-1$ of $k-m$ rest neighbors fail and transfer loads to it. Therefore, the probability that node $v$ comes to be an $m$-absorbing node with probability

\begin{equation}
\begin{split}
\label{eq:m_absorbing_node}
a_k^{(m)} = &\binom{k}{m} \sigma_a^m (1-\sigma_a)^{k-m} P\left\{L_v+(k-m-1)\Delta<(1+\alpha)L_v\right\} \\
= &\binom{k}{m} \sigma_a^m (1-\sigma_a)^{k-m} \left(1-\varphi\left(\frac{(k-m-1)\Delta}{\alpha}\right)\right) \; .
\end{split}
\end{equation}

We substitute Eq. (\ref{eq:m_absorbing_node}) into Eq. (\ref{eq:absorb_prob_decomposition}), and get the probability that node $v$ (i.e., arbitrary node of degree $k$) happens to be an absorbing node as

\begin{equation}
\label{eq:ak}
\begin{split}
a_k = &\sigma_a^k + \sum_{m=0}^{k-1}\binom{k}{m} \sigma_a^m (1-\sigma_a)^{k-m} \left( 1- \varphi\left( \frac{(k-m-1)\Delta}{\alpha}\right)\right) \\
=&
\begin{cases}
\sigma_a^k + \sum_{m=0}^{k-1}\binom{k}{m} \sigma_a^m (1-\sigma_a)^{k-m} \left( 1- \frac{(k-m-1)\Delta}{\alpha}\right) & \text{if $k \leq \left[ \alpha / \Delta\right]+1$} \\
\sigma_a^k + \sum_{m=k-1-\left[ \alpha / \Delta\right]}^{k-1}\binom{k}{m} \sigma_a^m (1-\sigma_a)^{k-m} \left( 1- \frac{(k-m-1)\Delta}{\alpha}\right) & \text{elsewhere}
\end{cases}
\\
=&
\begin{cases}
1-\frac{(k-1)\Delta}{\alpha}+\frac{k\sigma_a\Delta}{\alpha}-\frac{\sigma_a^k\Delta}{\alpha} & \text{if $k \leq \left[ \alpha / \Delta\right]+1$} \\
\sigma_a^k + \sum_{m=k-1-\left[ \alpha / \Delta\right]}^{k-1}\binom{k}{m} \sigma_a^m (1-\sigma_a)^{k-m} \left( 1- \frac{(k-m-1)\Delta}{\alpha}\right) & \text{elsewhere}
\end{cases}                 \; .
\end{split}
\end{equation}

According to the \textit{cascade condition} \cite{newman2001random,watts2002simple}, the absorbing nodes can not connect to each other to form a large connected sub-network with the following condition: 

\begin{equation}
\label{eq:ba_alpha_l_cond2}
\frac{1}{\langle k \rangle} \sum_{k=0}^{\infty} k(k-1)p_k a_k < 1  \; .
\end{equation}

Now we complete the derivation by Eqs. (\ref{eq:alpha_l_cond1}), (\ref{eq:sigma_a}), (\ref{eq:ak}), and (\ref{eq:ba_alpha_l_cond2}). These four equations depend on each other. We can get the threshold $\alpha_c$ with the following iteration process:

\begin{enumerate}
\item Assign $\left( \langle k \rangle - 1 \right) \Delta$ as an initial value to $\alpha$ according to Eq. (\ref{eq:alpha_l_cond1}).
\item \label{step:get_a_k} Get $a_k$ via Eq. (\ref{eq:sigma_a}) and (\ref{eq:ak}) with the current value of $\alpha$.
\item Substitute $a_k$ into Eq. (\ref{eq:ba_alpha_l_cond2}). If the inequality of Eq. (\ref{eq:ba_alpha_l_cond2}) is invalid, discount a very small value from $\alpha$, and then repeat step \ref{step:get_a_k}; otherwise,  let $\alpha_c = \alpha$ and finish the iteration process.
\end{enumerate}

\section{Simulation}
We validate the analytical prediction of threshold $\alpha_c$ in ER random and BA scale-free networks with $N = 5000$ nodes and varied average degree $\langle k \rangle = 8, 10, 12$. The load of each node is uniformly distributed on the interval $[0, 1]$. We randomly set very few nodes to fail as the beginning of the dynamical procedure. When a node fails, it transfers a fixed load $\Delta = 0.01$ to each of its un-failed neighbors. Fig. 3 and Fig. 4 present the giant component $G$ with different $\alpha$, and the threshold $\alpha_c$ for $G$ rises from zero to nonzero. Note that different values of $\Delta$ only changes the scale of abscissa. For each sub-figure in Fig. 3 and Fig. 4, each of 50 independent simulation results is draw respectively. The threshold $\alpha_c$ is marked by the solid triangle. It can be seen that our theoretical predictions are in good agreement with the simulation results.

\vspace*{4mm}

\begin{figure}
\setcounter{subfigure}{0}
\centering
\subfigure[$\langle k \rangle = 8$]{
\resizebox{0.31\textwidth}{!}{
  \includegraphics{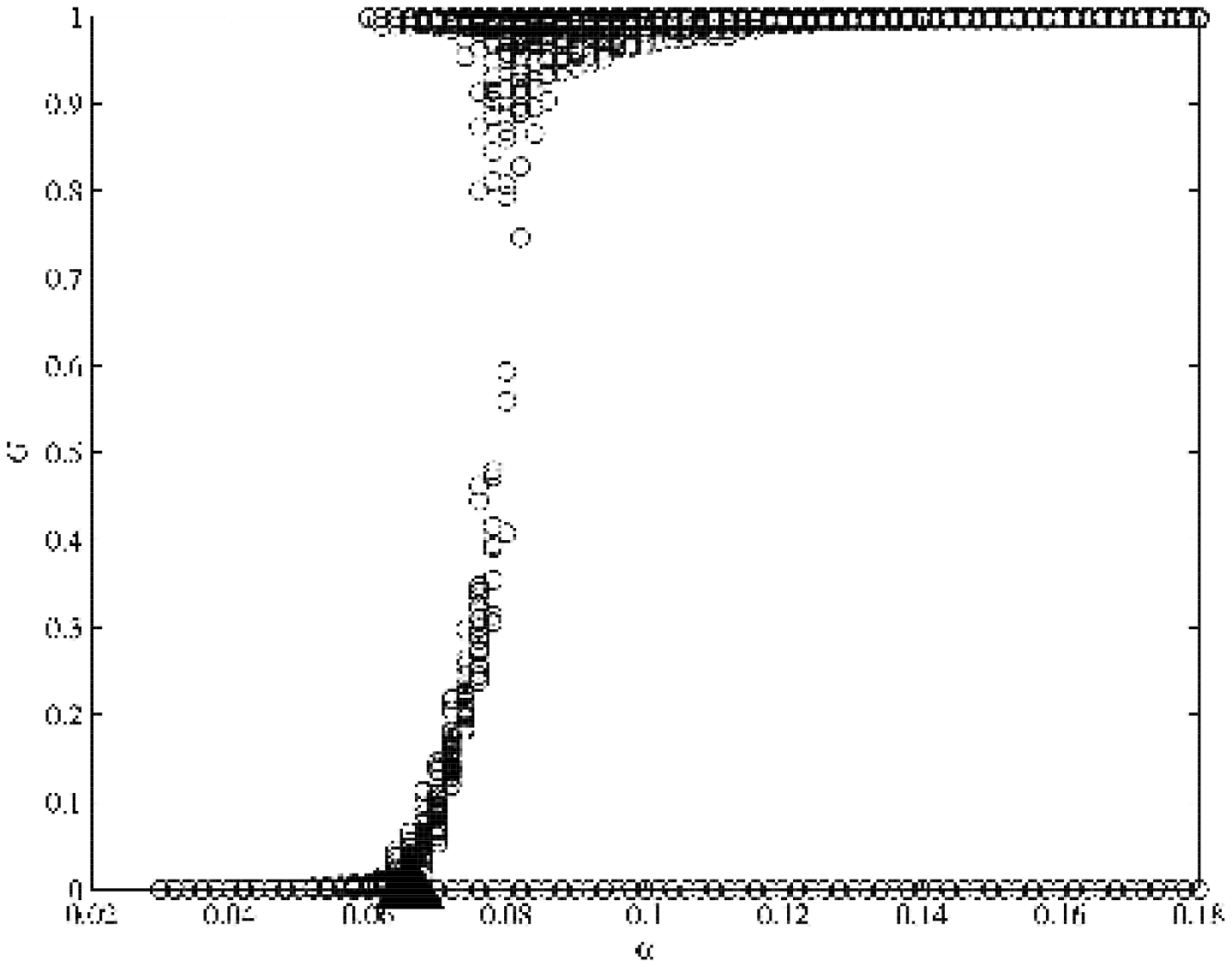}
}
}
\subfigure[$\langle k \rangle = 10$]{
\resizebox{0.31\textwidth}{!}{
  \includegraphics{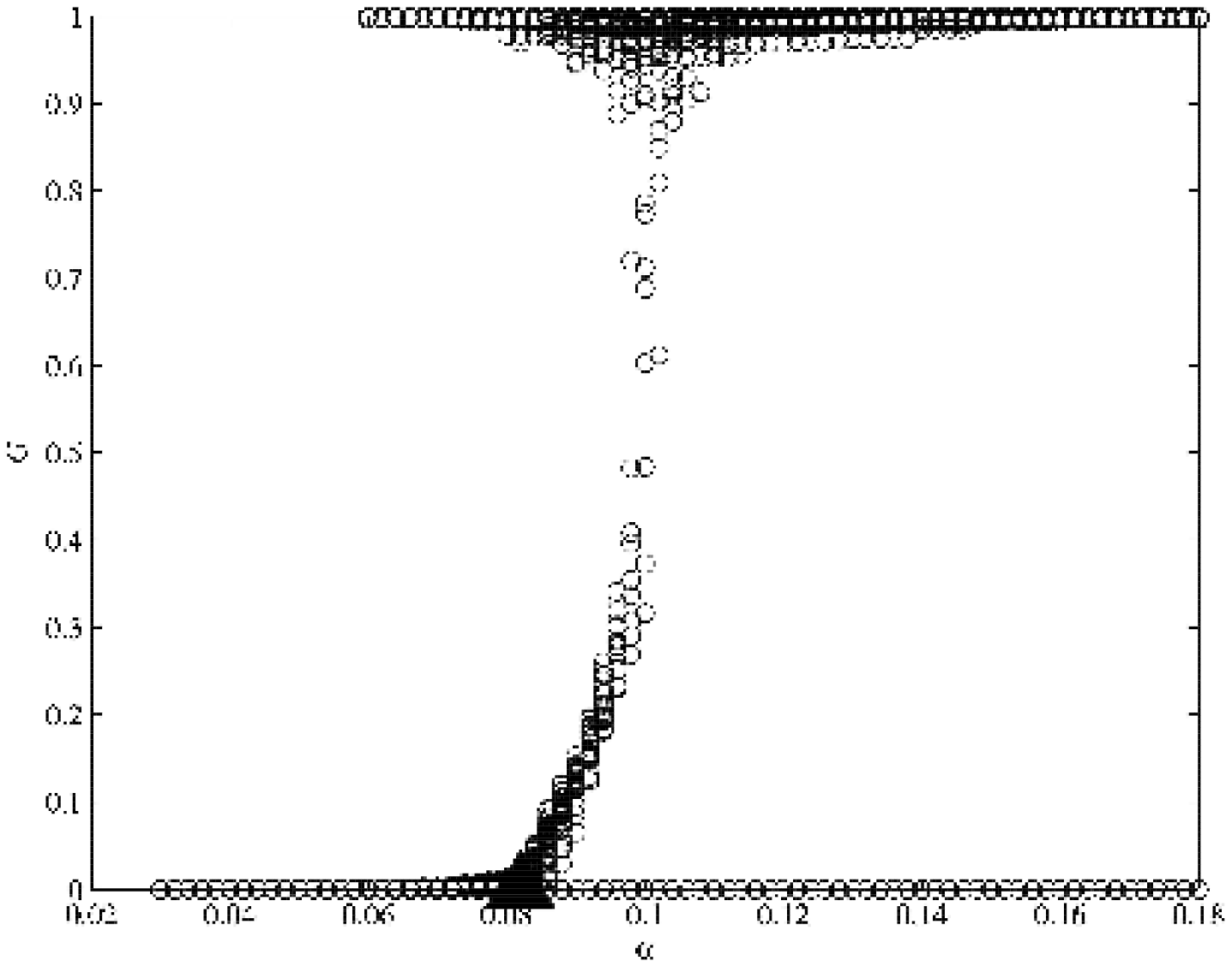}
}
}
\subfigure[$\langle k \rangle = 12$]{
\resizebox{0.31\textwidth}{!}{
  \includegraphics{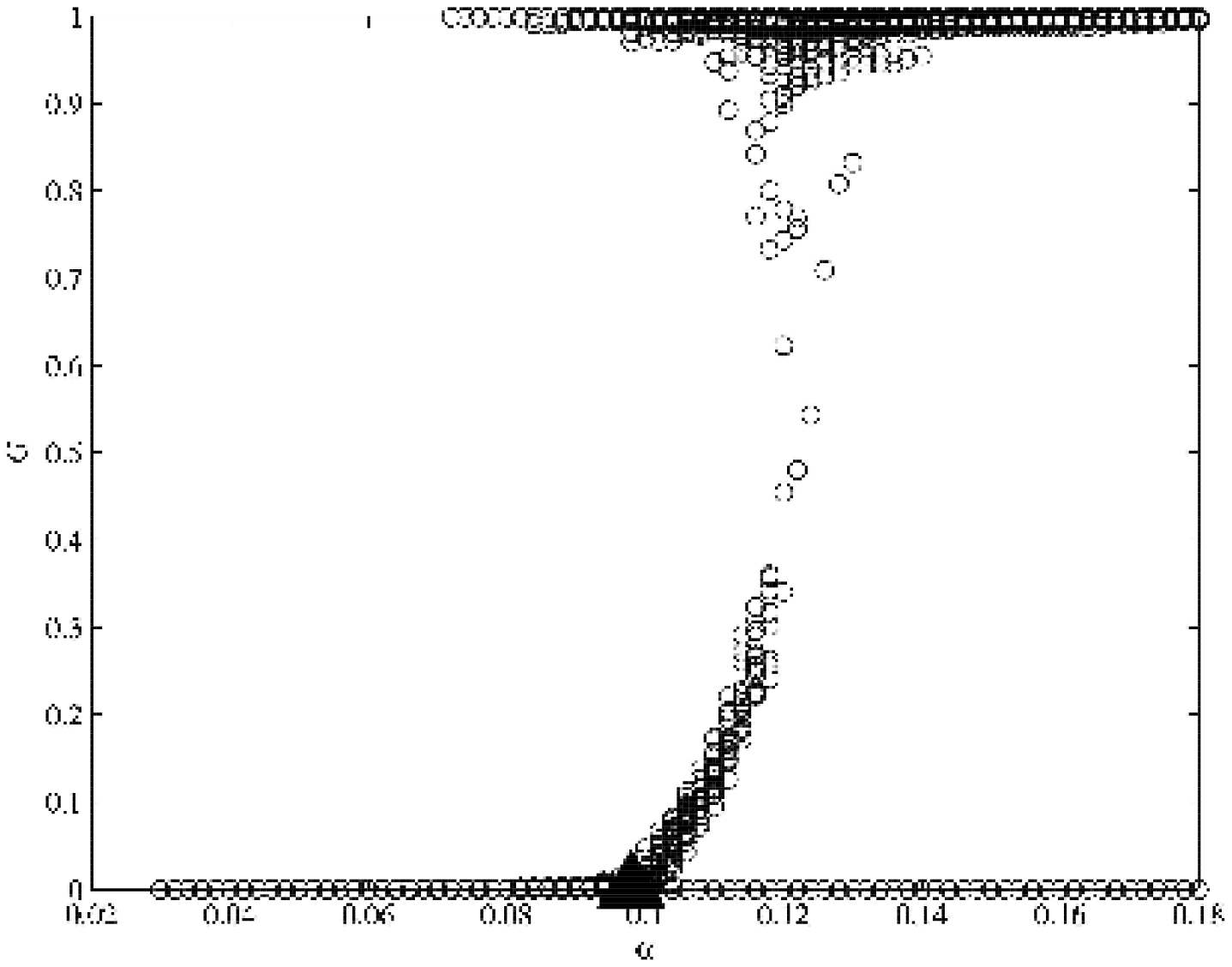}
}
}
\parbox{15.5cm}{\small{\bf Fig. 3}
Validation of the analytical predictions of threshold $\alpha_c$ where the giant component $G$ exists in ER random networks. With increase of tolerance parameter $\alpha$, we draw each simulated $G$ of 50 realizations respectively. The threshold $\alpha_c$ is marked by solid triangle ($\blacktriangle$).
}
\label{fig:er_cf_verification}
\end{figure}

\vspace*{4mm}

\begin{figure}
\setcounter{subfigure}{0}
\centering
\subfigure[$\langle k \rangle = 8$]{
\resizebox{0.31\textwidth}{!}{
  \includegraphics{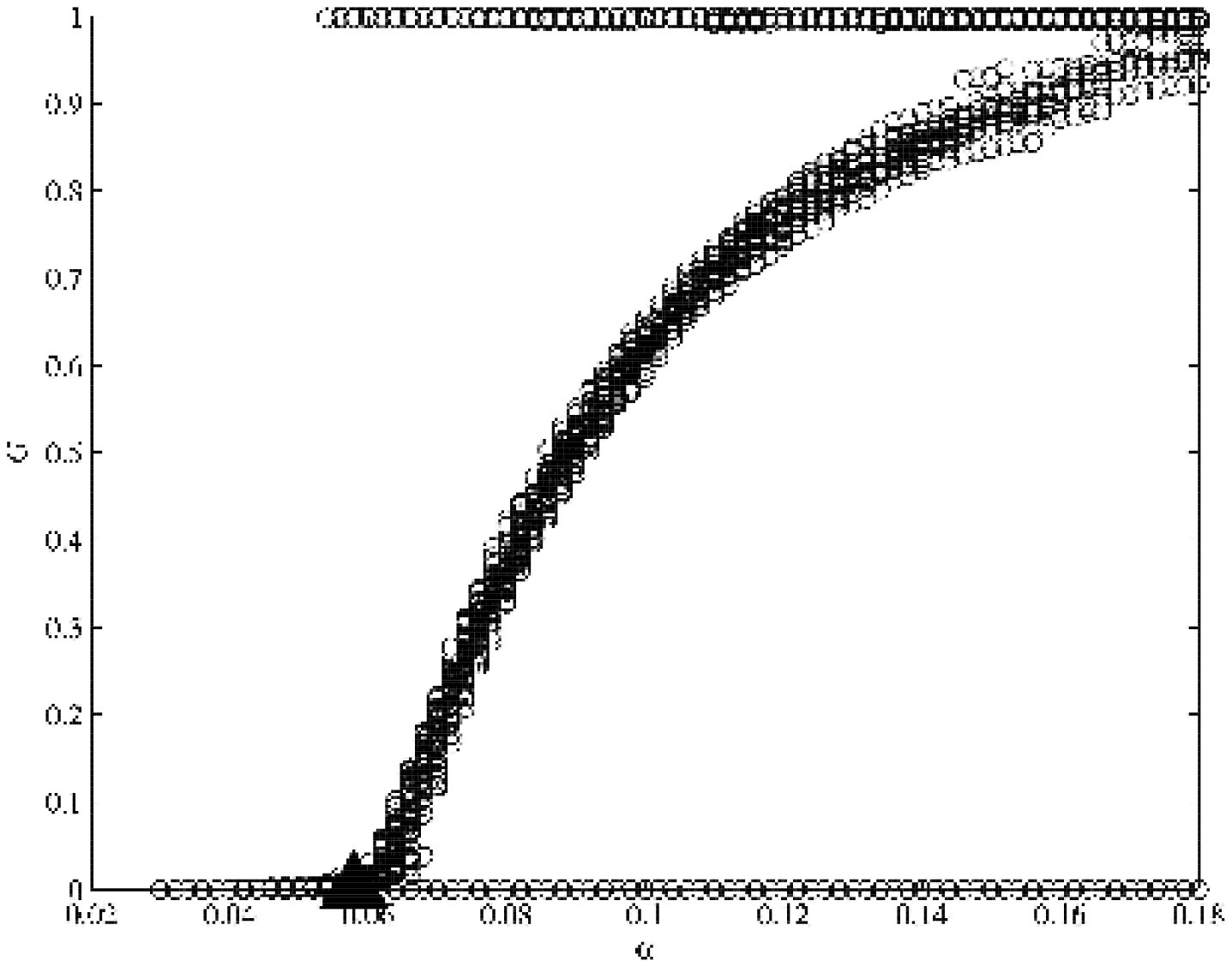}
}
}
\subfigure[$\langle k \rangle = 10$]{
\resizebox{0.31\textwidth}{!}{
  \includegraphics{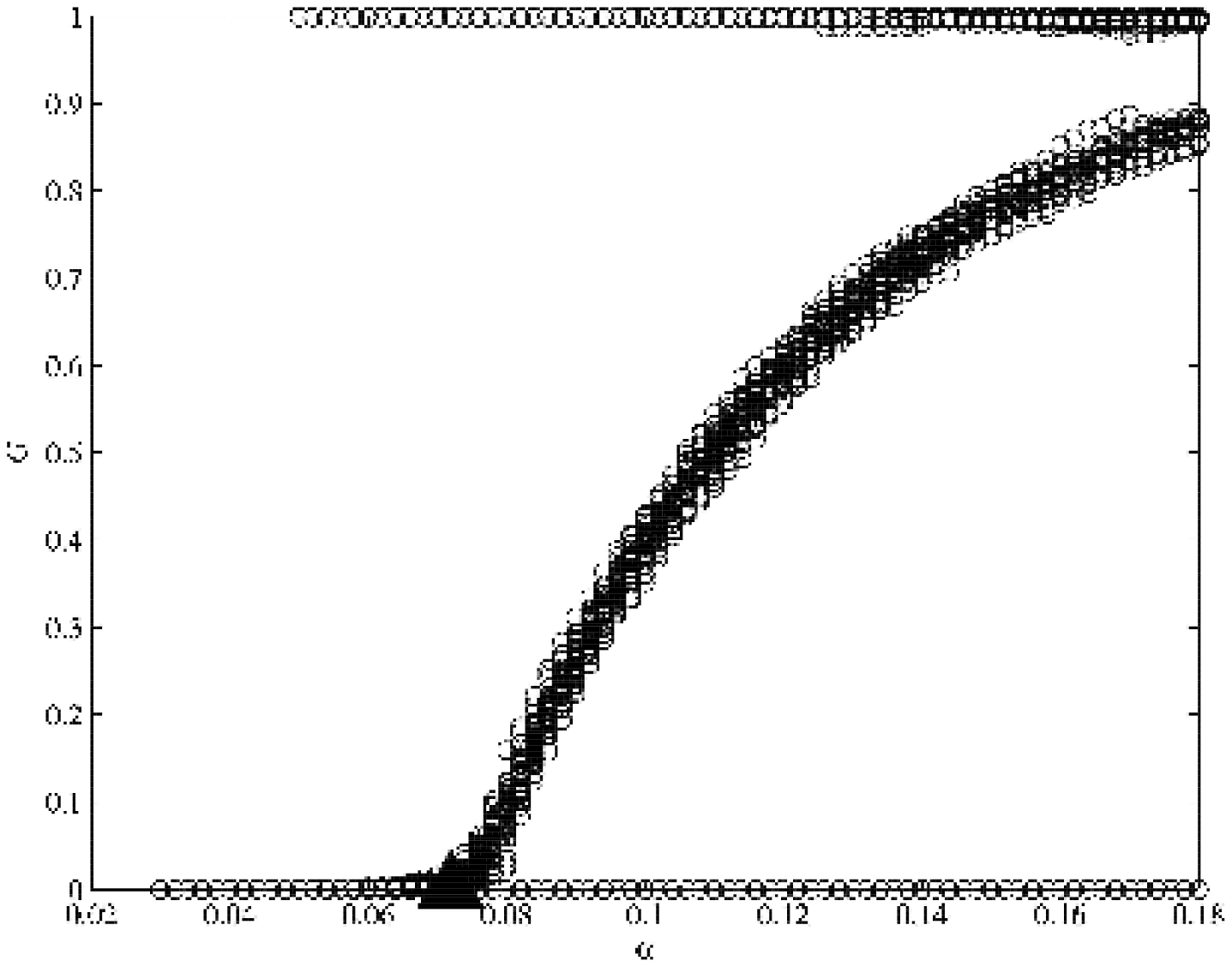}
}
}
\subfigure[$\langle k \rangle = 12$]{
\resizebox{0.31\textwidth}{!}{
  \includegraphics{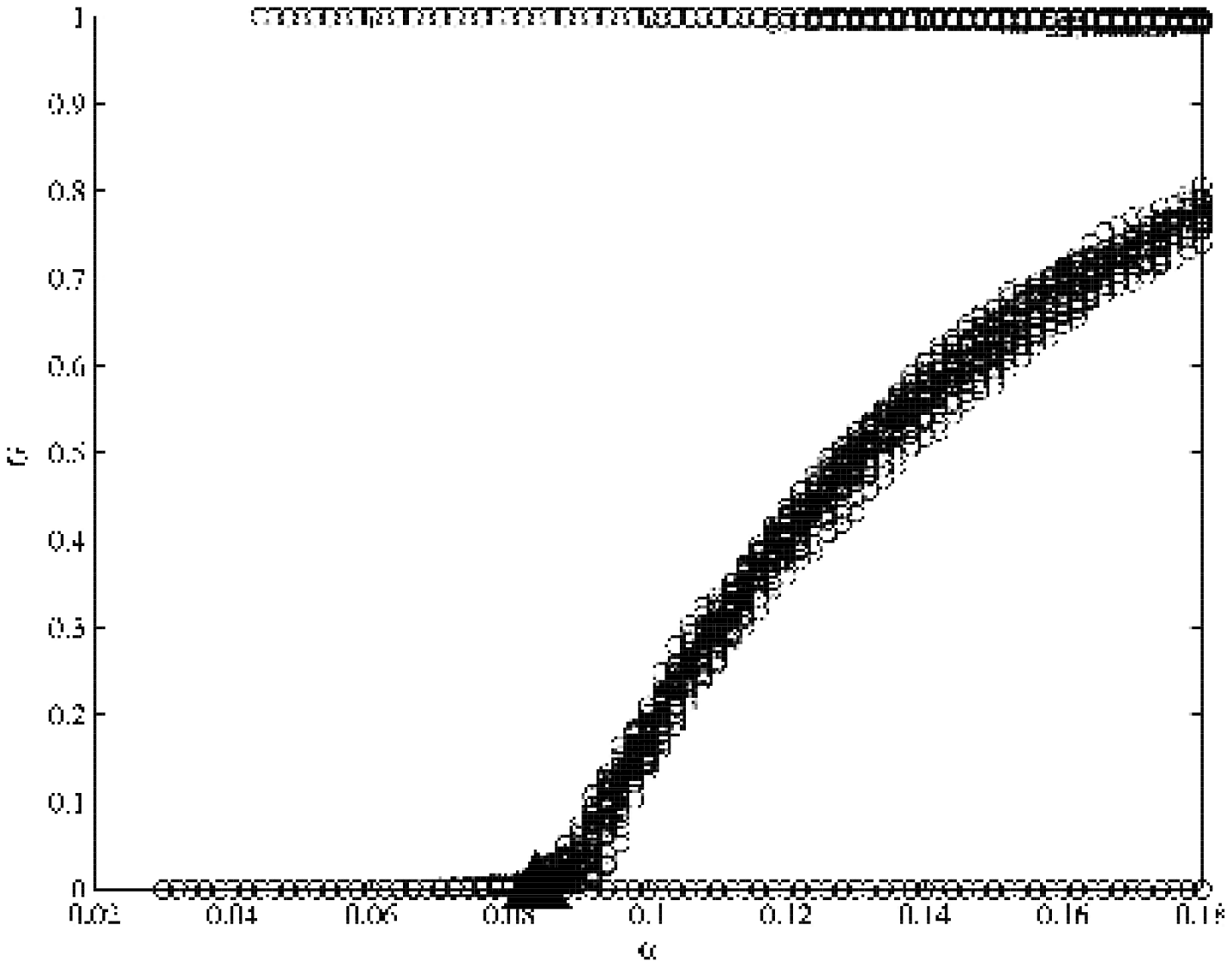}
}
}
\parbox{15.5cm}{\small{\bf Fig. 4}
Validation of the analytical predictions of threshold $\alpha_c$ where the giant component $G$ exists in BA scale-free networks. With increase of tolerance parameter $\alpha$, we draw each simulated $G$ of 50 realizations respectively. The threshold $\alpha_c$ is marked by solid triangle ($\blacktriangle$).
}
\label{fig:ba_cf_verification}
\end{figure}

\section{Conclusion}

In this paper, we provide a cascade of overload failure model with local load sharing mechanism, and then explore the threshold $\alpha_c$ of tolerance parameter for capacity, when the giant component $G$ formed by un-failed nodes rises from zero to non-zero. We provide two conditions to ensure the $G$ approximately equals to zero when $\alpha < \alpha_c$, which are: I. The large-scale cascading failures occur. II. un-failed nodes in steady state cannot connect to each other to form a large connected sub-network. With these two conditions, we get the theoretical derivation of $\alpha_c$ in degree-degree uncorrelated networks, and validate the effectiveness of this theoretical derivation in simulations. We believe that when creating a network and assigning load with local load sharing mechanism, this threshold $\alpha_c$ provides us a guidance to improve the network robustness under the premise of limited capacity resource.


\end{CJK*}  
\end{document}